\documentclass[twocolumn]{aastex631}
\usepackage[acronym]{glossaries}
\usepackage{hyperref}

\newacronym{gw}{GW}{gravitational wave}
\newacronym{grb}{GRB}{gamma ray burst}
\newacronym{ns}{NS}{neutron star}
\newacronym{bh}{BH}{black hole}
\newacronym{bns}{BNS}{binary neutron star}
\newacronym{bbh}{BBH}{binary black hole}
\newacronym{em}{EM}{electromagnetic}
\newacronym[plural=AGNs, firstplural=active galactic nuclei]{agn}{AGN}{active galactic nucleus}
\newacronym{sfft}{\texttt{SFFT}}{\texttt{Saccadic Fast Fourier Transform}}
\newacronym{lvk}{LVK}{LIGO/Virgo/KAGRA}
\newacronym{o4}{O4}{the fourth gravitational wave observing run}
\newacronym{o5}{O5}{the fifth gravitational wave observing run}
\newacronym{psc}{PSC}{Pittsburgh Supercomputing Center}
\newacronym{nersc}{NERSC}{National Energy Research Scientific Computing Center}
\newacronym{decam}{DECam}{Dark Energy Camera}
\newacronym{cp}{CP}{DECam Community Pipeline}
\newacronym{des}{DES}{Dark Energy Survey}
\newacronym{desi}{DESI}{Dark Energy Spectroscopic Instrument}
\newacronym{decals}{DECaLS}{DECam Legacy Survey}
\newacronym{delve}{DELVE}{DECam Local Volume Exploration survey}
\newacronym{ls}{LS}{DESI Legacy Survey}
\newacronym{mpc}{MPC}{Minor Planet Center}
\newacronym{tns}{TNS}{Transient Name Server}
\newacronym{cbc}{CBC}{compact binary coalescence}
\newacronym{gcn}{GCN}{General Coordinates Network}
\newacronym[plural=KNe, firstplural=kilonovae]{kn}{KN}{kilonova}
\newacronym{too}{ToO}{Target of Opportunity}
\newacronym{healpix}{HEALPix}{Hierarchical Equal Area isoLatitude Pixelation}
\newacronym{gwmmads}{GW-MMADS}{Gravitational Wave MultiMessenger Astronomy DECam Survey}
\newacronym{snr}{SNR}{signal-to-noise ratio}
\newacronym{noirlab}{NOIRLab}{NSF National Optical-Infrared Research Laboratory}
\newacronym{cnn}{CNN}{convolutional neural network}
\newacronym{parsnip}{ParSNIP}{Parameterization of SuperNova Intrinsic Properties}
\newacronym[plural=SNe, firstplural=supernovae]{sn}{SN}{supernova}
\newacronym{blr}{BLR}{broad-line region}
\newacronym{wise}{WISE}{Wide-field Infrared Survey Explorer}
\newacronym{sf}{SF}{structure function}
\newacronym{ned}{NED}{NASA/IPAC Extragalactic Database}
\newacronym{tde}{TDE}{tidal distruption event}
\newacronym{bpt}{BPT}{Baldwin–Phillips–Terlevich}
\newacronym{lsst}{LSST}{Legacy Survey of Space and Time}
\newacronym{ztf}{ZTF}{Zwicky Transient Facility}
\newacronym[plural=SLSNe, firstplural=superluminous supernovae]{slsn}{SLSN}{superluminous supernova}
\newacronym{bhl}{BHL}{Bondi-Hoyle-Littleton}
\newacronym{smbh}{SMBH}{supermassive black hole}
\newacronym{ctio}{CTIO}{Cerro Tololo Inter-American Observatory}
\newacronym{ir}{IR}{infrared}
\newacronym{ci}{CI}{confidence interval}
\newacronym{lris}{LRIS}{Low-Resolution Imaging Spectrograph}
\newacronym{gmos}{GMOS}{Gemini Multi-Object Spectrograph}
\newacronym{dragons}{DRAGONS}{Data Reduction for Astronomy from Gemini Observatory North and South}
\newacronym{salt}{SALT}{South African Large Telescope}
\newacronym{rss}{RSS}{Robert Stobie Spectrograph}
\newacronym{pc}{PC}{Photon Counting}
\newacronym{3kk}{3KK}{Three Channel Camera}
\newacronym{ps1}{PS1}{PanSTARRS 1}
\newacronym{2mass}{2MASS}{Two Micron All Sky Survey}
\newacronym{nir}{NIR}{near-infrared}

\begin{document}

\title{A GPU-Accelerated Transient Detection Pipeline for DECam Time-Domain Surveys}

\newcommand{\mcwilliams}{
    McWilliams Center for Cosmology and Astrophysics,
    Department of Physics,
    Carnegie Mellon University,
    5000 Forbes Avenue, Pittsburgh, PA 15213
}

\newcommand{\tamu}{
    George P. and Cynthia Woods Mitchell Institute for Fundamental Physics \& Astronomy, \\
    Texas A. \& M. University, Department of Physics and Astronomy, 4242 TAMU, College Station, TX 77843, USA
}

\newcommand{\tamids}{Texas A\&M Institute of Data Science
John R. Blocker Building, Suite 227
155 Ireland Street, TAMU 3156
College Station, TX 77843-3156}

\newcommand{\unc}{
    Department of Physics and Astronomy, 
    University of North Carolina at Chapel Hill, 
    Chapel Hill, NC 27599-3255, USA
}

\correspondingauthor{Lei Hu}
\email{leihu@sas.upenn.edu}

\author[0000-0001-7201-1938]{Lei Hu}
\affiliation{\mcwilliams}
\affiliation{Department of Physics and Astronomy, University of Pennsylvania, Philadelphia, 209 South 33rd Street, PA 19104, USA}

\author[0000-0002-1270-7666]{Tom\'as Cabrera}
\affiliation{\mcwilliams}

\author[0000-0002-6011-0530]{Antonella Palmese}
\affiliation{\mcwilliams}

\author[0000-0001-7092-9374]{Lifan Wang}
\affiliation{\tamu}

\author[0000-0002-8977-1498]{Igor Andreoni}
\affiliation{\unc}

\author[0000-0002-9364-5419]{Xander J. Hall}
\affiliation{\mcwilliams}

\author[0000-0003-3021-4897]{Xingzhuo Chen}
\affiliation{\tamu}
\affiliation{\tamids}

\author[0000-0002-1376-0987]{Jiawen Yang}
\affiliation{\tamu}

\author[0000-0001-5567-1301]{Frank Valdes}
\affiliation{NSF NOIRLab, 950 N Cherry Ave, Tucson, AZ 85719, USA}

\author[0000-0002-9700-0036]{Brendan O'Connor}
\affiliation{\mcwilliams}

\author[0009-0007-2247-2315]{Yuhan Chen}
\affiliation{Department of Physics, University of California, Berkeley, CA 94720, USA}






\begin{abstract}

We present a GPU-accelerated transient detection pipeline developed for time-domain surveys with the Dark Energy Camera (DECam). It enables real-time-capable image processing, incorporating science-driven candidate filtering to support rapid transient identification in time-critical observing programs. 
The pipeline serves as the core transient discovery engine for multiple long-term DECam programs, including the GW-MMADS gravitational-wave follow-up campaign and the DESIRT survey for intermediate-redshift transients with DESI synergy. 
The pipeline ingests calibrated imaging products from the DECam Community Pipeline and performs image differencing using the SFFT algorithm, coupled with CNN-based real-bogus classification, to produce science-ready transient alerts and light curves that are delivered to community brokers.
We validate the pipeline using archival DECam data from the DESIRT survey. The real-bogus classifier achieves a completeness of $\sim$ 99\% of real transients while rejecting $\sim$ 96\% of subtraction artifacts, and the workflow typically reduces the candidate load to a manageable level for survey operations. 
With GPU acceleration, the typical processing time per DECam exposure is $\sim$ 50 s from calibrated image processing to alert generation using a modest allocation of computing resources.

\end{abstract}



\section{Introduction} \label{sec:intro}

Dark Energy Camera \citep[DECam;][]{DECAM2015} is a wide-field optical imager mounted on the 4-m Victor M. Blanco Telescope at the Cerro Tololo Inter-American Observatory in Chile. Originally designed for the Dark Energy Survey \citep[DES;][]{DES2016}, DECam comprises 62 science CCDs, delivering a field of view of approximately 3~deg$^2$ with a pixel scale of 0.263~arcsec/pixel. With the completion of the Dark Energy Survey in 2019, DECam has continued to enable a broad range of time-domain opportunities through dedicated transient surveys \citep[e.g.,][]{DECamHITS2016,DECamDWF2019,DECamYSE2022,DECamDDF2023,DESIRT_2022,DESGW2020,S230922g_Tomas2024}.

We have developed a GPU-accelerated transient-detection pipeline to support heterogeneous DECam time-domain observing programs. The pipeline is designed to operate robustly across diverse observing modes, spanning event-driven Target-of-Opportunity (ToO) observations and regularly scheduled survey observations.
Its primary objective is to deliver science-ready transient candidates with minimal program-specific tuning and rapid turnaround, facilitating efficient downstream analysis and timely follow-up observations. 

At its core, our pipeline is built upon image differencing for transient detection. Difference image analysis (DIA) has long served as a transient discovery engine in time-domain astronomy \citep{AL98,Bramich2008,Miller2008,HOTPANTS,ZOGY,SFFT2022}.
In this work, we employ the GPU-accelerated Saccadic Fast Fourier Transform algorithm\footnote{\url{https://github.com/thomasvrussell/sfft}} (SFFT; \citealt{SFFT2022}) for image differencing. 
The \textsc{SFFT} algorithm formulates the image subtraction problem in the Fourier domain and accounts for spatial variations in the point-spread function (PSF), which are ubiquitous in wide-field astronomical imaging such as DECam. 
The convolution kernel is modeled with a $\delta$-function basis \citep{Bramich2008,Miller2008}, providing enhanced flexibility, accommodating subpixel astrometric misalignments, and requiring minimal user-adjustable parameters.
Notably, the Fourier-domain formulation enables efficient and scalable computation via fast Fourier transforms, thereby supporting GPU acceleration and high-throughput processing in large-scale time-domain surveys.
Related efforts to incorporate GPU-accelerated SFFT into transient-detection and photometry pipelines for time-domain applications on other wide-field facilities include, for example, \citet{WFST_Pipeline_Cai2025} and \citet{Phrosty2025}.

The DECam pipeline presented here supports multiple long-term DECam time-domain programs.
These include GW-MMADS (Gravitational Wave Multi-Messenger Astronomy DECam Survey; PIs: Andreoni $\&$ Palmese), a DECam-based gravitational-wave follow-up campaign conducted throughout the LIGO/Virgo/KAGRA O4 observing run, enabling systematic electromagnetic counterpart searches for a diverse range of gravitational-wave events, such as S230922g \citep{S230922g_Tomas2024}, S250206dm \citep{S250206dm_Hu2025}, and S250818k \citep{S250818k_Hall2025}.

The pipeline has also been in sustained use by the DESIRT survey \citep[DECam Intermediate-Redshift Transient Survey; PIs: Palmese $\&$ Wang;][]{DESIRT_2022} since 2021. Unlike GW-MMADS that operates in an event-driven ToO mode, DESIRT is an ongoing DECam time-domain program characterized by a typical cadence of approximately three days and multi-band (\textit{griz}) imaging over $\sim$100~deg$^2$, reaching depths of $r \sim 23.5$~mag. The survey is designed to discover and characterize transients at intermediate redshifts ($0.1 < z < 0.3$) in coordination with the DESI spectroscopic observations. The initial DESIRT campaign ran from early 2021 to late 2023, followed by a successor program with a similar observing strategy that began in early 2025, the DECam DESI Transient Survey (PI: Palmese; \citealt{2DTS_Hall2025, DESI_Transients_Hall2026}).

In addition, the pipeline supports an ongoing six-semester DECam survey\footnote{\protect\url{https://time-allocation.noirlab.edu/\#/proposal/details/589422}} (PI: D. Sand) aimed at identifying the youngest nearby supernovae by ``shadowing'' LSST Wide-Fast-Deep observations of nearby galaxy structures. 
It has further been applied to other time-critical programs such as the DECam+ZTF fast transient search experiment \citep{2024TNSAN..68....1A}, targeting rapidly evolving transients through DECam observations in tandem with ZTF.


This paper is structured as follows. In Section~\ref{sec:pipeline}, we present a comprehensive description of the DECam transient detection pipeline, detailing the end-to-end workflow from pre-observation preparation to science-ready products. In Section~\ref{sec:example_pipeline_products}, we demonstrate its application using DECam observations from the DESIRT survey and present representative candidate filtering results and associated pipeline outputs. In Section~\ref{sec:pipeline_performance}, we evaluate the pipeline performance, including artifact rejection and computational efficiency. 

\section{Pipeline} \label{sec:pipeline}

\begin{figure*}[ht!]
\plotone{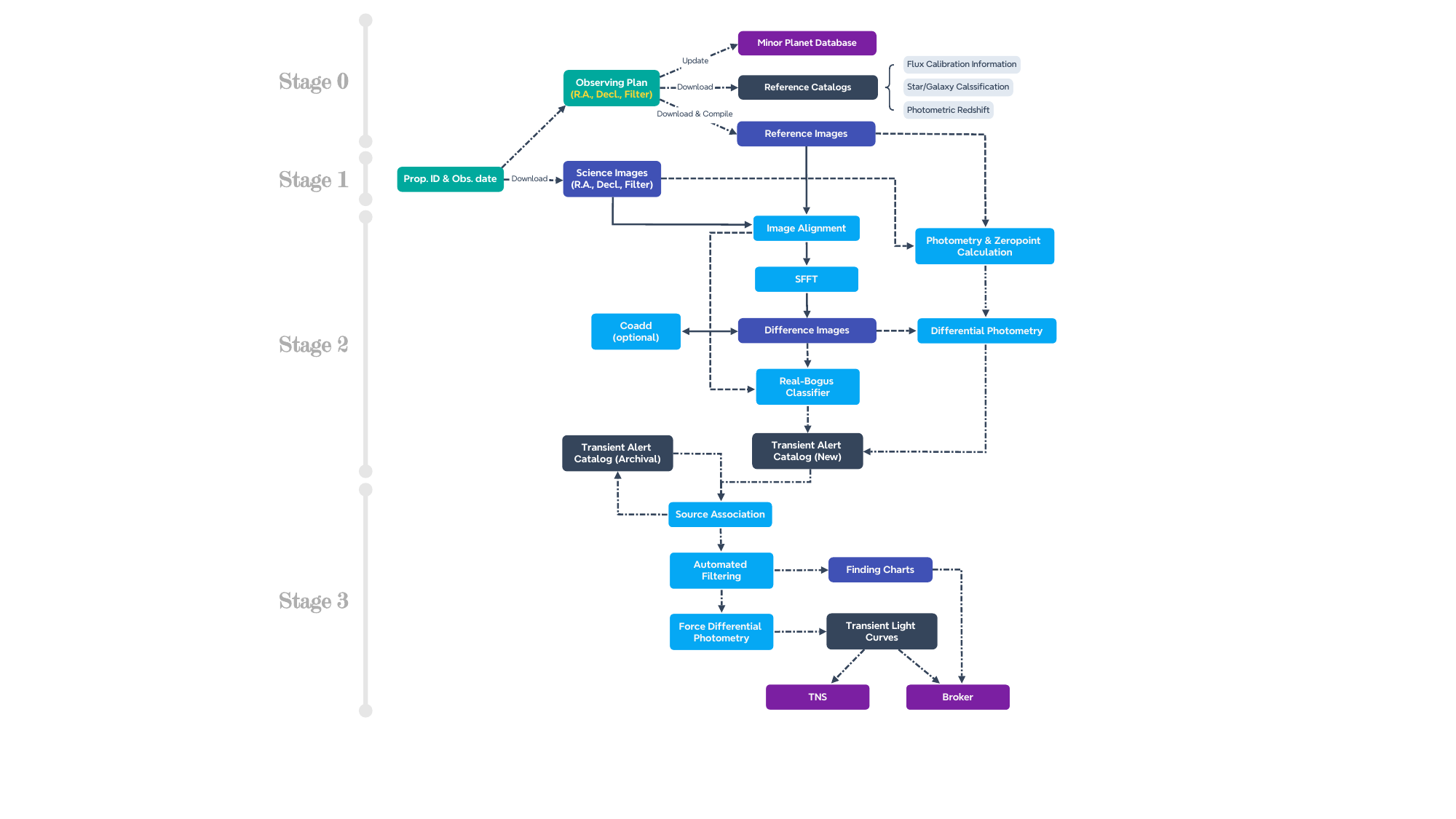}
\caption{Schematic overview of the DECam transient detection pipeline. Colored boxes denote pipeline inputs, intermediate data products, processing components, and internal/external services. Arrows indicate data flow between modules, with solid arrows representing paths involving imaging data. The pipeline is divided into four stages, spanning pre-observation preparation (Stage~0), science data ingestion (Stage~1), image differencing and transient detection (Stage~2), and candidate filtering and science product generation (Stage~3). \label{fig:workflow}}
\end{figure*}

In this section, we describe the design of our DECam transient-detection pipeline. The pipeline is organized into four stages: pre-observation preparation (Stage~0; Section~\ref{subsubsec:prepsteps}), science data ingestion (Stage~1; Section~\ref{subsubsec:stage1}), image differencing and transient detection (Stage~2; Section~\ref{subsubsec:stage2}), and candidate filtering with science product generation
(Stage~3; Section~\ref{subsubsec:stage3}). Figure~\ref{fig:workflow} shows a schematic overview of the pipeline workflow.

\subsection{Stage 0: preparatory steps before observations}\label{subsubsec:prepsteps}

This stage includes all preparatory steps that can be completed before new observations become available.
It is particularly useful for time-critical programs such as gravitational-wave follow-up campaigns (e.g., GW-MMADS), where rapid downstream processing is required immediately after data acquisition.
The primary goal of this stage is to prepare an auxiliary dataset, including template images, external catalogs, and metadata queries needed for real-time calibration, differencing, and transient detection.

\textit{\bf DECam Observing plan} — The pipeline is initialized with a DECam observing plan specifying the scheduled sky coordinates and filter for each exposure, with required columns \texttt{R.A.}, \texttt{Decl.}, and \texttt{Filter} \footnote{In typical operation, the observing plan is supplied by the user prior to observations. If the observations have already been executed, the corresponding observing records retrieved from the NOIRLab archive may also be used as the observing plan.}.
Each plan is associated with a unique program identifier \texttt{PROGRAM\_ID} and an observing date \texttt{CALDAT}.
For DECam, the instrument footprint is effectively fixed for a given pointing, subject to typical pointing uncertainties generally no greater than 20 arcseconds. 

We do not require specific sky field names to be provided in the observing plan. Instead, each pointing is assigned a pipeline-native field identifier based on the HEALPix tessellation scheme. We adopt \texttt{NSIDE=$2^{12}$}, corresponding to a spatial resolution of approximately $50''$ per cell (or $\sim200$ DECam pixels), which safely exceeds expected pointing uncertainties. The pipeline-native field name (e.g., \texttt{HP195286455}) is used to ensure that identical or highly overlapping pointings consistently access the same auxiliary datasets (template images, catalogs, etc.) across an observing campaign. For simplicity, we refer to each planned pointing (identified by a unique HEALPix index) together with its DECam filter as a \textit{scheduled instance} in the following sections.

\textit{\bf NOIRLab archival metadata mirror} — A local PostgreSQL relational database is established to maintain a synchronized mirror of DECam observation metadata from the NOIRLab Astro Data Archive\footnote{\url{https://astroarchive.noirlab.edu/}} for all available DECam archival images. 
The metadata mirror stores essential exposure information and observing characteristics to facilitate accelerated interaction with the Data Archive and support our template selection, and also records direct download URLs to streamline programmatic retrieval of both template and science images.

\textit{\bf Prepare galaxy catalog} — We also use the PostgreSQL database to manage an amalgamated galaxy catalog for later host identification.
The pipeline initializes this catalog with the NED-LVS \citep{NEDLVS_2023} by default\footnote{The catalog was expanded to include galaxies from the DESI survey in a later pipeline upgrade.}.

\textit{\bf Minor Planet Center (MPC) database} — We also maintain a local, regularly synchronized copy of the Minor Planet Center (MPC) orbit database\footnote{\url{https://minorplanetcenter.net/iau/MPCORB.html}}, which is used for automated filtering of known Solar System objects during transient candidate vetting in Stage~3 (Section~\ref{subsubsec:stage3}).

\subsubsection{Template image preparation} \label{subsubsec:temp_prep}

For each \textit{scheduled instance} specified in the observing plan, our pipeline supports two options for template image construction: single-exposure template and multi-exposure templates. 
Single-exposure template is suitable for programs in which the new observations are intentionally designed to match existing archival pointings, and where the archival exposures are typically deeper than the new data. This is the typical case for programs such as GW-MMADS and DESIRT. Multi-exposure templates are adopted when one or more of these conditions are not satisfied. We describe the two modes in detail below.

Throughout this paper, all archival data retrieved from the NOIRLab archive are calibrated products produced by the DECam Community Pipeline (CP; \citealt{CP_2014}) and accessed via the NOIRLab archival metadata mirror.
Specifically, we use the calibrated science images (\texttt{ooi}) together with their corresponding data quality masks (\texttt{ood}), which flag known bad pixels to facilitate artifact rejection in Stage~2 (Section~\ref{subsubsec:stage2}).

\textit{\bf Single-exposure template} — For this mode, we query the NOIRLab archival metadata mirror to identify the best-matched single archival exposure to serve as the template image.
Candidate templates are required to satisfy the following criteria:
(1) the exposure is taken with the same filter and its pointing center lies within $2.18'$ (or $\sim500$ DECam pixels) of the scheduled position;
(2) the observation date is at least 30 days prior to the science exposure (a tunable parameter in the pipeline).
Among all candidates, the optimal template is selected as the one with the largest effective exposure time $t_{\mathrm{eff}}$. The effective exposure time, as defined in \citet{osti_1250877}, is proportional to the nominal exposure time and inversely related to the ratio of the squared FWHM and sky background relative to a reference condition (FWHM = $0.9''$ and dark sky at zenith).
Ties in $t_{\mathrm{eff}}$ are broken in preference of the exposure with the longer exposure time.
The selected archival exposure is retrieved for the \textit{scheduled instance} using the stored download URL recorded in the metadata mirror.

\begin{table*}[!t]
\centering
\caption{Auxiliary catalogs used in the pipeline.}
\label{tab:aux_catalogs}
\begin{tabular}{llll}
\hline
Reference Catalog name & Role & Query scope & Reference \\
\hline
Legacy Survey DR10
& Flux calibration; Star/galaxy separation 
& {\it Scheduled instance} 
& \citet{LS2019} \\
Pan-STARRS1 DR1
& Flux calibration
& {\it Scheduled instance}
& \citet{PS1Survey} \\
\textit{Gaia} DR3 
& Flux calibration
& {\it Scheduled instance}
& \citet{GAIA_DR3} \\
NED-LVS 
& Galaxy redshift (spectroscopic)
& Local full copy
& \citet{NEDLVS_2023} \\
Legacy Survey Sweeps
& Galaxy redshift (photometric)
& {\it Scheduled instance}
& \citet{LS2019} \\
\hline
\end{tabular}
\tablecomments{For each type of use, catalogs are listed in order of descending priority from top to bottom.}
\end{table*}

\textit{\bf Multi-exposure templates} — This mode is used when either the new observations are not well aligned with existing archival pointings, or when the best available archival data are not deep enough for sensitive transient detection.

For each CCD detector of a given \textit{scheduled instance}, we identify candidate archival exposures by evaluating the overlap between each archival DECam image and the target CCD footprint.
An archival exposure is selected as a candidate if its overlap area exceeds 80\% of the CCD detector.
Among all candidates, we rank exposures by their effective exposure time $t_{\mathrm{eff}}$ and select the top \texttt{N\_TEMP} (a tunable parameter with a default of 5) as the template set for that CCD detector.

Each selected archival exposure is then resampled onto the expected WCS of the target CCD detector to produce a synthetic CCD image. Repeating this procedure for all CCD detectors yields a synthetic template exposure.
We can therefore construct \texttt{N\_TEMP} synthetic template exposures for each \textit{scheduled instance}. By design, a single synthetic template exposure may be constructed from different archival images, whereas all synthetic templates are mapped onto a common WCS.

Coaddition of these synthetic templates would require careful handling of PSF matching to avoid spatial discontinuities introduced by CCD gaps, which are not well modeled by image differencing and may degrade subtraction quality. For simplicity, we retain these outputs as a set of independent template exposures without coaddition.

The data quality masks associated with the selected archival exposures undergo the same resampling procedure as the images. Since only binary information (good versus bad pixels) is required, the resampled \texttt{ood} images are discretized using an empirical threshold of 0.3: pixels with values above 0.3 are flagged as bad (set to $2^1$), and others are flagged as good (set to $2^0$).

\subsubsection{Auxiliary catalog preparation} \label{subsubsec:auxcat_prep}

Auxiliary catalogs are used in our pipeline for three primary purposes:
(1) photometric flux calibration,
(2) star--galaxy separation for preliminary classification of detected variable sources,
and (3) host-galaxy redshift assignment.
The main reference catalogs used for these purposes are summarized in Table~\ref{tab:aux_catalogs}.
For each scheduled instance, the pipeline constructs a dedicated set of auxiliary catalogs
by querying multiple external surveys with complementary roles, as listed in Table~\ref{tab:aux_catalogs}.

The DESI Legacy Imaging Survey \citep{LS2019} serves as the primary reference catalog for photometric calibration and star--galaxy separation through its morphological {\tt TYPE} classification.
The catalog is divided into spatially indexed \textit{tractor bricks} across the sky. For each scheduled DECam pointing, all \textit{tractor bricks} that overlap the footprint are retrieved and merged into a locally cached reference catalog for downstream processing. 
When Legacy Surveys coverage is unavailable for a given pointing and filter, Pan-STARRS1 (PS1; \citealt{PS1Survey}) photometry is used as a secondary reference, followed by \textit{Gaia} DR3 \citep{GAIA_DR3} as the final fallback.
For PS1 and \textit{Gaia}, cone searches are performed via \texttt{astroquery.vizier}
to retrieve all catalog entries within a region that fully covers each scheduled DECam footprint.

To place all reference photometry onto the DECam photometric system, PS1 photometry is transformed into the DECam/DES system using the color transformation equations adopted in DES DR2 \citep{DES_DR2_2021}, mapping PS1 $grizy$ magnitudes to the DES $grizY$ system for sources within the recommended color ranges.
In addition, \textit{Gaia} DR3 photometry is converted into the DECam/DES system following the prescription of \citet{GAIA2DES}, which provides transformations from Gaia magnitudes to DECam $g$, $r$, $i$, and $z$.

Host-galaxy redshifts are preferentially adopted from a locally mirrored NED-LVS catalog
\citep{NEDLVS_2023} whenever spectroscopic measurements are available.
When no spectroscopic redshift is found, photometric redshifts from the Legacy Surveys sweeps catalogs%
\footnote{\url{https://www.legacysurvey.org/dr10/photoz/}} are used as a supplement.

\subsection{Stage 1: Ingest new observations} \label{subsubsec:stage1}

Once the CP-calibrated products of new observations become available from the NOIRLab archive\footnote{The turnaround time is typically within the same night, ranging from tens of minutes to several hours depending on the time-critical priority of the program.}, the pipeline automatically ingests each exposure by downloading the calibrated science image (\texttt{ooi}) and its associated data-quality mask (\texttt{ood}) via the URLs stored in the local archival metadata mirror.

At this stage, we also construct a standardized directory hierarchy to organize all data products required for downstream image processing.
Science images and their derived products are organized hierarchically by program name, \textit{scheduled instance} (defined by the pointing HEALPix index and observing filter), observing night, and individual science exposure identifiers
\footnote{\mbox{For example, \texttt{S250206dm/HP194008839/DECam-r/2025-02-12/}} \texttt{c4d\_250213\_074045\_xxx\_r\_gw} for an individual science exposure.}.

Template images and auxiliary reference catalogs are placed at a higher-level directory and symbolically linked at the program and \textit{scheduled instance} level\footnote{\mbox{For example, \texttt{S250206dm/HP194008839/DECam-r/TEMPLATE/}} and \mbox{\texttt{S250206dm/HP194008839/DECam-r/AUXILIARY}} denote the directories for the
template images and auxiliary reference catalogs, respectively.}, such that all science exposures within the same field and filter share a consistent set of reference products.

\subsection{Stage 2: Image processing and transient detection} \label{subsubsec:stage2}

\subsubsection{Photometric calibration} \label{subsubsec:zero_point}

Aperture photometry is performed on the CCD images of each exposure (both science and template) using \textsc{SExtractor} \citep{SExtractor}, with point-like sources cross-matched against external reference catalogs (Section~\ref{subsubsec:auxcat_prep}) to determine the photometric zero point associated with the adopted aperture.

For single-exposure templates, the aperture radius is defined as a fixed multiple of the FWHM, with a default value of 0.6731 times the FWHM, the optimal size for sources whose point-spread functions can be approximated as Gaussian \citep{Faherty_2014}.
For multi-exposure templates, we instead adopt a fixed aperture of 3 pixels in radius, corresponding approximately to the Gaussian-optimal size for the median PSF FWHM of typical DECam exposures.

\subsubsection{Image differencing} \label{subsubsec:img_diff}

We first reproject the new science image onto the WCS of the corresponding template image(s) using \textsc{SWarp} \citep{SWarp}, ensuring pixel-level alignment between the science and template frames.
In the case of multi-exposure templates (Section~\ref{subsubsec:temp_prep}), all templates share a common WCS, and thus the science image is resampled only once.

We perform rapid image differencing for each science image using the GPU-accelerated Saccadic Fast Fourier Transform\footnote{\url{https://github.com/thomasvrussell/sfft}} (SFFT; \citealt{SFFT2022}).
We adopt the default software configuration throughout this work, and find that the method robustly delivers high-quality difference images for DECam over a wide range of observing conditions.
For multi-exposure templates, the science image is differenced against each individual template exposure, producing {\tt N\_TEMP} difference images per science exposure. All difference images are subsequently normalized to a common photometric zero point (Section~\ref{subsubsec:zero_point}) and median-combined to produce a single final difference image for downstream analysis.

\subsubsection{Differential Photometry} \label{subsubsec:diff_photometry}

We perform source detection on the difference images using \textsc{SExtractor}, adopting the same photometric aperture and zero-point as those used for the photometric calibration (Section~\ref{subsubsec:zero_point}) of the unconvolved image in the image-differencing step (Section~\ref{subsubsec:img_diff}).
Spurious detections contaminated by bad pixels flagged in the CP data quality masks ({\tt ood}) of either the science or template images are removed. This procedure yields an initial set of candidate transient and variable sources for each DECam exposure.

At this step, we also estimate the limiting magnitude of each difference image via a Monte Carlo approach. We measure aperture fluxes at 1024 randomly selected background positions using the same aperture definition, and deriving the 5$\sigma$ threshold from their standard deviation, converted to magnitudes with the corresponding zero point (Section~\ref{subsubsec:zero_point}).

\subsubsection{Real-bogus classification} \label{subsubsec:real_bogus}

In this step, each differential detection in the differential photometry catalog (Section~\ref{subsubsec:diff_photometry}) is assigned a real-bogus score to distinguish real astrophysical signals from subtraction artifacts.
This scoring is performed using a modified rotation-invariant convolutional neural network (CNN) classifier originally proposed by \citet{sunPipelineAntarcticSurvey2022}.
The model is a hybrid architecture based on the rotation-invariant framework of \citet{DeepHiTSRotationInvariant}, incorporating residual connections inspired by ResNet-18 \citep{heDeepResidualLearning2015}.
For each candidate detected in the difference image (Section~\ref{subsubsec:diff_photometry}), the classifier takes as input a set of three-channel image stamps—consisting of the science image, template image, and their difference—centered on the candidate position, and outputs a scalar probability score ranging from 0 (artifact) to 1 (real).

The classifier is trained using archival DECam imaging data, augmented with simulated transient source injections to improve the representation of real astrophysical signals.
We summarize the construction of the classifier as follows:
\begin{itemize}
    \item \textbf{Dataset:} We use DESIRT DECam imaging data obtained during the 2021A semester as the dataset for classifier training and evaluation. We selected observations acquired over a two-month period (April--May 2021), covering 19 observing nights with a typical cadence of approximately three days. In total, the dataset includes 1,661 exposures in the $g$, $r$, and $z$ bands, which are sufficient to be representative of the DECam data processed by our pipeline.
    
    \item \textbf{Transient injection:} The two-month subset of DESIRT observations provides a limited number of real transient sources for training a CNN-based classifier. We therefore augment the dataset by injecting simulated transient point sources into the DESIRT 2021A imaging data.
    For each DECam CCD image in the dataset, we randomly select up to 50 extended sources classified as galaxies in the Legacy Survey catalog \citep{LS2019} and inject simulated point sources in their vicinity, with injection positions randomly drawn within a radius of 4\arcsec\ from the galaxy center and source magnitudes uniformly sampled between 17.0~mag and the $5\sigma$ limiting magnitude of the CCD image.
    These transient injections are implemented with the AstrOmatic software suite, with PSF models constructed using \textsc{SExtractor} \citep{SExtractor} and \textsc{PSFEx} \citep{PSFEx2011}, and point sources simulated using \textsc{SkyMaker} \citep{SkyMaker2009}.

    \item \textbf{Sample construction and labeling:} Following transient injection, we run the pipeline on the DECam dataset with simulations to generate difference images (Section~\ref{subsubsec:img_diff}) and differential photometry catalogs (Section~\ref{subsubsec:diff_photometry}). 
    All injected sources that are successfully re-detected in the difference images are labeled as real, yielding 137,232 samples. The remaining detections are labeled through visual inspection. Any detection not identified as a subtraction artifact is classified as real, including variable stars, AGN-like events, and minor planets. In total, the final labeled sample contains 198,802 real detections and 157,428 artifact detections.

    \item \textbf{Model training:} The labeled dataset is randomly split into training, validation, and test subsets with an 80\%/5\%/15\% ratio, and the classifier used for our pipeline is trained accordingly. The performance of the trained model evaluated on the test set is presented in Section~\ref{subsec:rejection_performance}.
\end{itemize}

\subsubsection{Alert generation} \label{subsubsec:alert_gen}

The differential photometry catalog is further filtered by applying a real--bogus score threshold of ${\tt CNN\_SCORE\_THRESH}$, yielding a transient alert catalog for each DECam exposure. 
By default, we adopt a hot threshold for the CNN real–bogus classifier, corresponding to ${\tt CNN\_SCORE\_THRESH}=0.3$, which favors sensitivity over purity. This choice is particularly well suited for applications where recovering faint transients is critical. For example, in gravitational-wave follow-up programs, the hot threshold is used to enhance sensitivity to potential faint kilonova counterparts.
The pipeline can also be configured to use a cold, purity-favoring threshold (e.g., ${\tt CNN\_SCORE\_THRESH}=0.6$), which prioritizes high-confidence detections and substantially reduces the number of spurious candidates passed to downstream inspection. The performance of the CNN classifier under different threshold choices is discussed in Section~\ref{subsec:rejection_performance}.

We annotate each alert with the following external astronomical context based on its sky position and observation time. 
\begin{itemize}
    \item \textbf{Minor planet association:} Each alert is cross-matched with known minor planets by propagating ephemerides from the MPC database (Section~\ref{subsubsec:prepsteps}) to the observation time using the \textsc{ephem} package. The angular separation to the nearest predicted minor planet position is recorded as an auxiliary metric for subsequent candidate vetting; 
    
    \item \textbf{Stellar source association:} We identify the nearest point-like source in the Legacy Survey catalog \citep{LS2019}, classified as ${\tt TYPE = PSF}$. The angular separation to the matched source ({\tt SEP\_ALT2STAR}, in arcsec) is used to flag alerts that are likely associated with stellar variability.
    
    \item \textbf{Host galaxy association:} We use the directional light radius (DLR) method \citep{Sullivan_2006,Gupta_2016,Qu_2024,DES_2024} to associate each alert with its potential host galaxy in the Legacy Survey catalog \citep{LS2019}. 
    Host candidates are restricted to extended sources within 0.1$^\circ$ of the alert position, identified by morphological classifications (i.e., ${\tt TYPE} \in \{{\tt REX}, {\tt DEV}, {\tt EXP}, {\tt SER}\}$).

    To reduce spurious associations with faint or unreliable Legacy Survey detections, additional photometric quality constraints are applied to the host candidate sample. Specifically, candidate hosts are required to satisfy at least two of the following magnitude criteria: $g < 24.0$, $r < 23.4$, and $z < 22.5$, and to have colors within $-1 < g-r < 4$.

    For each alert, the angular separation to the matched host galaxy ({\tt SEP\_ALT2GAL}, in arcsec) is recorded as part of the alert metadata.
\end{itemize}

\subsection{Stage 3: Candidate filtering and science product generation} \label{subsubsec:stage3}

\subsubsection{Source association} \label{subsubsec:source_assoc}

This step assigns a unique object identifier by associating alerts from
different exposures into a single object based on sky position. Alerts from all DECam exposures in the program are processed in chronological order of observation time.  
For each alert, we search for an existing object within a matching radius of 1.9$^{\prime\prime}$. If no such object is found, a new object ID is created; otherwise, the alert is associated with the nearest existing object. The chronological processing of alerts ensures that object identifiers assigned in earlier nights remain unchanged during subsequent nightly processing, enabling incremental and reproducible object association.

Each object identifier encodes its preliminary classification, discovery date, and sky position, determined from the first alert associated with the object. As an example, the identifier {\tt C202504181415355m003413} is constructed as follows. 

The leading character denotes a preliminary object type based on the alert position relative to nearby cataloged sources: {\tt V} (variable-like), {\tt A} (AGN-like), {\tt C} (transient near a galaxy core), or {\tt T} (transient off the galaxy core), according to the angular separations defined in Section~\ref{subsubsec:alert_gen}.
An object is classified as {\tt V} if it is closely associated with a nearby stellar source, with ${\tt SEP\_ALT2STAR} < 2$~arcsec; otherwise, the classification is determined by the separation to the matched host galaxy, with thresholds ${\tt SEP\_ALT2GAL}\le0.3$~arcsec ({\tt A}), $0.3<{\tt SEP\_ALT2GAL}<1$~arcsec ({\tt C}), and ${\tt SEP\_ALT2GAL}\ge1$~arcsec ({\tt T}). The remaining characters encode the observing date and sky position of the alert, with {\tt p} and {\tt m} indicating positive and negative declination, respectively. 

\subsubsection{Candidate filtering} \label{subsubsec:candidate_filtering}

A set of selection criteria is applied to identify plausible transient candidates from the full object catalog generated in Section~\ref{subsubsec:source_assoc}.
The baseline cuts, which typically reject the majority of objects, are defined as follows:
\begin{itemize}
    \item {\bf Exclusion of stellar variability}: Objects with a preliminary type of {\tt V} are excluded. To account for residual stellar contaminants not fully captured by the morphology-based classification from the Legacy Survey, objects are further flagged as stellar if they are matched to point sources in the Gaia DR3 catalog \citep{GAIA_DR3} with measured proper motion exceeeding 1.081~mas/yr  at high significance \citep{GAIA_DR2_Astrometry_2018}.
    
    \item {\bf Exclusion of known minor planets}: Alerts associated with known minor planets within 10~arcsec are flagged and excluded from subsequent filtering. This criterion is applied at the alert level; objects containing additional alerts not associated with minor planets are retained.
\end{itemize}

The pipeline further supports additional user-configurable candidate filtering tailored to observing strategies and specific science goals. 
Examples of commonly adopted customized criteria include:
\begin{itemize}
    \item {\bf Multiple alert requirements}: Objects may be required to have at least ${\tt NALT\_THRESH}$ alerts, with the time separation between the earliest and latest alerts exceeding a configurable threshold, ${\tt ALT\_DTIME}$.
    Typically, ${\tt NALT\_THRESH}$ is set to 2 or 3, and ${\tt ALT\_DTIME}$ is chosen to range from a few hours to half a day; this configuration is found to be effective in filtering out residual artifacts and faint moving-object contamination not included in the MPC catalog.
    \item {\bf Discovery time constraints}: Objects can be selected based on the discovery time of their first alert, restricted to a specified time window. This is particuarly useful for identifying newly emerging transients.
\end{itemize}

\subsubsection{Science product generation} \label{subsubsec:science_product_gen}

The pipeline generates a set of user-facing science products for each selected candidate. An example illustrating these products for a representative object is presented in the following Section~\ref{sec:example_pipeline_products}.

\paragraph{Finding charts.} Finding charts consist of triplet image stamps (reference, science, and difference images) centered on the object position. For diagnostic purposes, the pipeline generates such triplet stamps for all science exposures that cover the object location and stores them internally. For user-facing products, a compact set of representative epochs is selected, including the last non-detection, the first detection (alert), and the detection (alert) with the highest signal-to-noise ratio (S/N).

\paragraph{Transient light curves.} While differential photometry is available for each difference image (Section~\ref{subsubsec:diff_photometry}), the final transient light curves are
constructed using forced aperture photometry on all difference images covering the object, with the aperture centered at a fixed sky position defined by the highest-S/N differential detection.
The adopted aperture sizes and photometric zero-points follow those described in Section~\ref{subsubsec:diff_photometry}. 
When possible, the transient light curves are additionally augmented with absolute magnitudes using the redshift of the associated host galaxy.
For objects whose first alert is associated with a Legacy Survey host with DLR ratio $<4$ (Section~\ref{subsubsec:alert_gen}), spectroscopic redshifts retrieved from our auxiliary NED-LVS catalog are adopted when available; otherwise, photometric redshifts from the Legacy Survey sweeps catalog are used (Section~\ref{subsubsec:auxcat_prep}).

At this stage, a final human-in-the-loop vetting step is typically performed based on the generated science products to identify a short list of the most promising transient candidates.
Candidates that pass this inspection are then programmatically disseminated
to the Transient Name Server\footnote{\url{https://www.wis-tns.org/}} (TNS) and the astronomical data platform \textsc{SkyPortal}\footnote{\url{https://skyportal.io/}}, enabling broad community access and
rapid coordination of follow-up observations \citep{walt_skyportal_2019}.

The pipeline also packages all Stage~3-selected candidates into ZTF-style Avro alerts. These alerts are currently compatible with the BOOM broker \citep{jegou_du_laz_boom_2025}, which enables rapid cross-matching of DECam alerts with those from other facilities.



\section{An example with DESIRT survey} \label{sec:example_pipeline_products}
\begin{deluxetable}{ccc}
    \tablecaption{
        Example candidate-filtering results from the DESIRT 2025A DECam transient search. \label{tab:vetting}
    }
    \tablehead{ & Number & Fraction \\
    Vetting Filter & passed & passed}
    \startdata
        \multicolumn{3}{c}{\textit{Regular pipeline cuts}} \\
        \hline
        Alert produced on difference image & 909,714 & 1.000 \\
        Not stellar variability & 702,421 & 0.772 \\
        Not MPC object & 248,456 & 0.273 \\
        \hline 
        \multicolumn{3}{c}{\textit{User-configurable pipeline cuts}} \\
        \hline
        Multiple detections & 7,646 & 0.008 \\
        Discovered after April 15, 2025 & 2,737 & 0.003 \\
        \hline
        \multicolumn{3}{c}{\textit{Human-in-the-loop cuts}} \\
        \hline
        Visual inspection & 76 & $8 \times 10^{-5}$ \\
    \enddata
    \tablecomments{The quantities listed in the table represent the number of objects that passed each step in the filtering process.}
\end{deluxetable}
 
\begin{figure}[ht!]
\includegraphics[width=1.0\columnwidth]{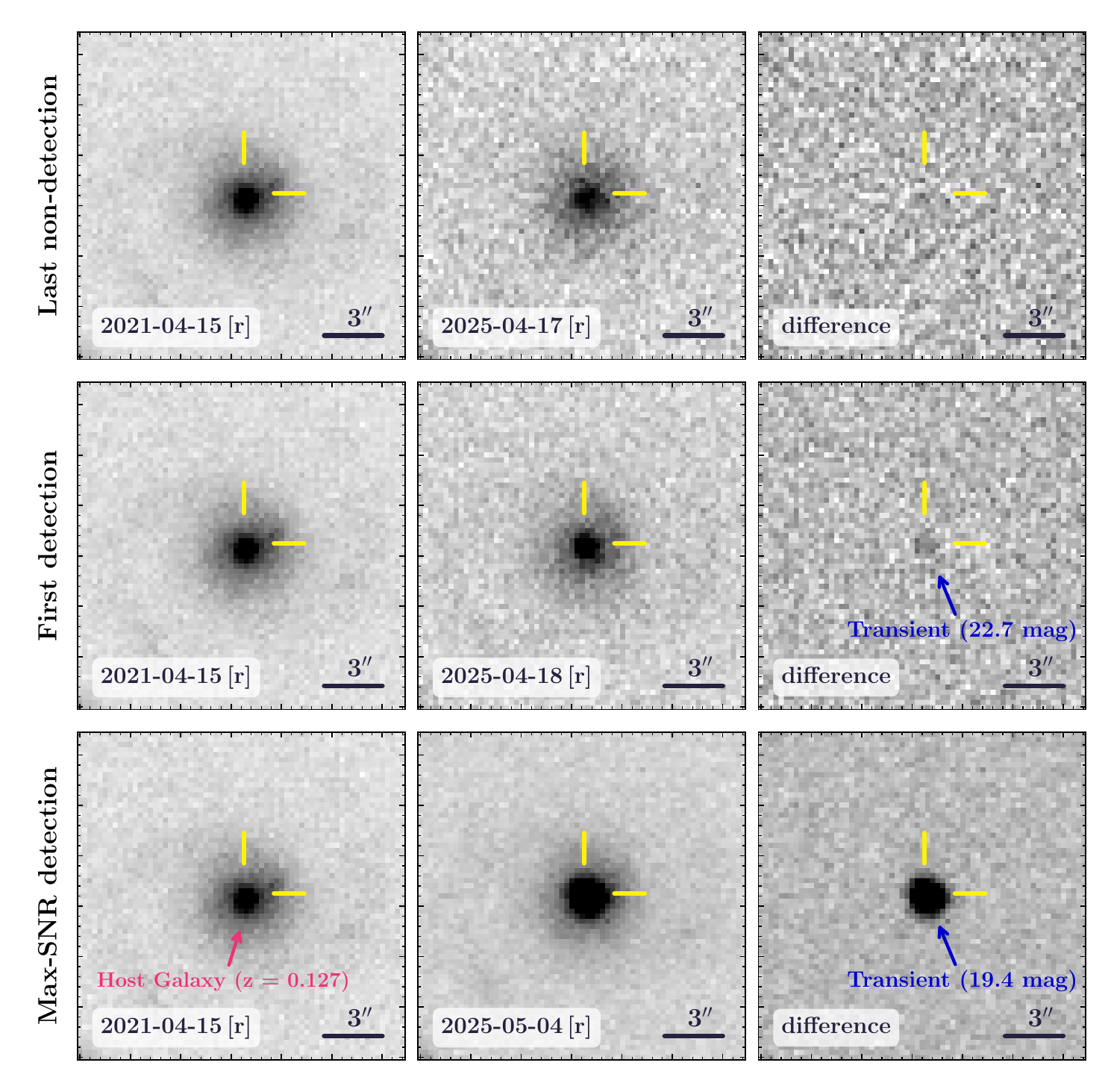}
\caption{
    Example pipeline finding chart product for the transient candidate AT~2025ifl. Each row shows image cutouts centered on the transient position at different epochs: the last non-detection (top), the first detection (middle), and the epoch with the maximum S/N (bottom). The transient position is indicated by yellow crosshairs, and detections on the difference images are highlighted with blue arrows and labeled with their apparent magnitudes. The host galaxy at $z=0.127$ \citep{NEDLVS_2023} is marked in red. \label{fig:pipeline_product_FC}
}
\end{figure}

\begin{figure*}[ht!]
\plotone{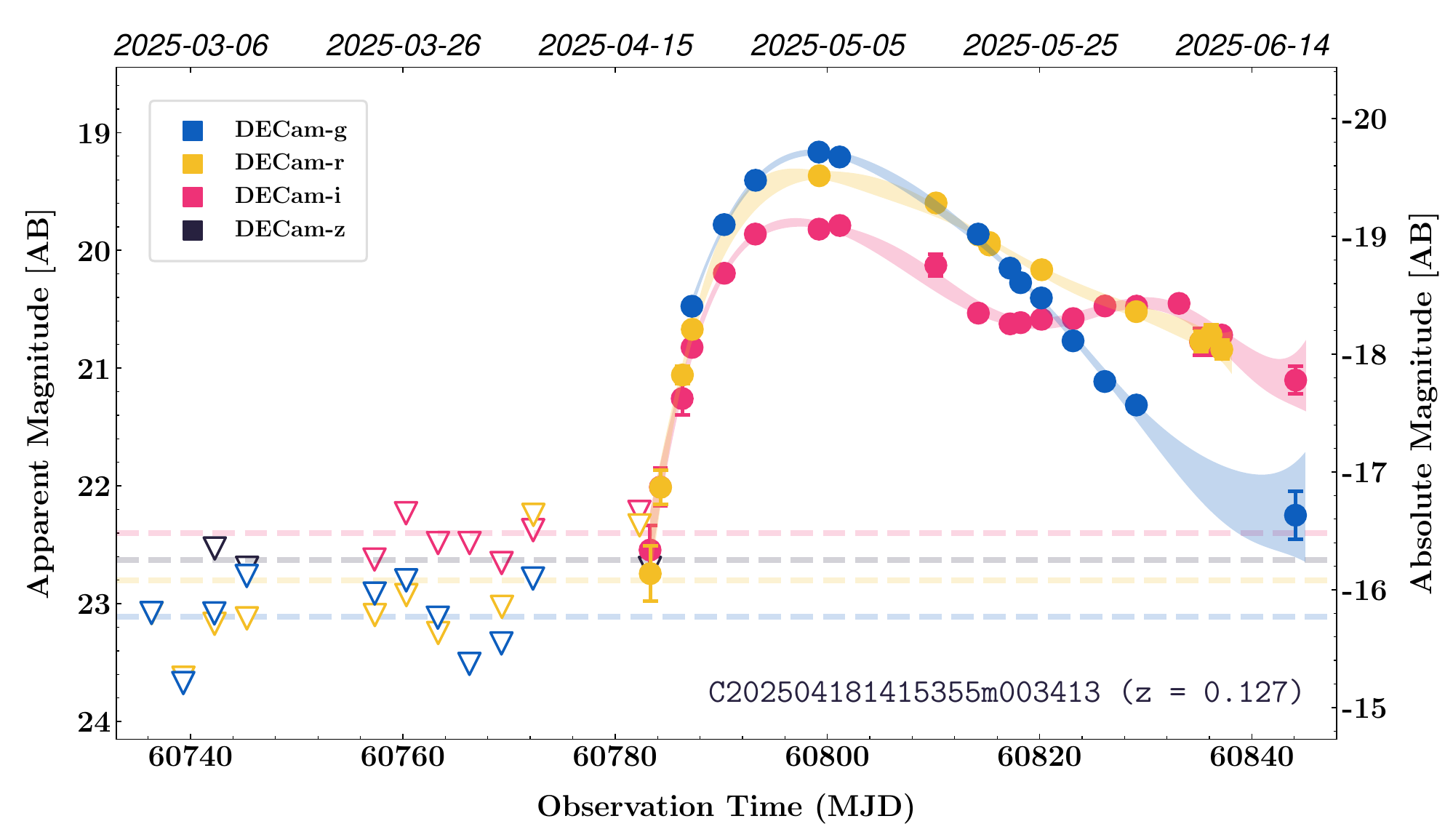}
\caption{
    Example pipeline light-curve product for the transient candidate AT~2025ifl. Multi-band DECam force photometry produced by the pipeline is shown as filled circles, while non-detections are indicated by downward triangles with dashed horizontal lines marking the average $5\sigma$ limiting magnitudes. The shaded regions indicate the $\pm2\sigma$ confidence intervals of the Gaussian process fits \citep{sklearn_2011} to the measured light curves in each band. The absolute magnitudes are computed using the spectroscopic redshift of the host galaxy at $z=0.127$ \citep{NEDLVS_2023}. \label{fig:pipeline_product_LC}}
\end{figure*}

In this section, we apply our pipeline described in Section~\ref{sec:pipeline} to the DECam data collected by the DESIRT survey during the 2025A semester to showcase example pipeline results.
The science dataset consists of DECam observations obtained between February and June 2025, spanning 55 observing nights and comprising 3,642 exposures in the $g$, $r$, $i$, and $z$ bands.

The pipeline is configured for a general transient search with the following settings.
For alert generation, we adopt by default a hot real--bogus score threshold of ${\tt CNN\_SCORE\_THRESH}=0.3$ (Section~\ref{subsubsec:alert_gen}).
In the candidate-filtering stage (Section~\ref{subsubsec:candidate_filtering}), a requirement of ${\tt NALT\_THRESH}=3$ alerts with a minimum temporal baseline of ${\tt ALT\_DTIME}=0.5$ day is imposed.
For illustration purposes, we restrict the candidate sample to the objects discovered after April 15, 2025, corresponding to the latter half of the dataset.

The candidate-filtering results at each stage are summarized in Table~\ref{tab:vetting}. Starting from a total of 909,714 objects produced from alerts on difference images (Section~\ref{subsubsec:source_assoc}), successive pipeline cuts reduce the sample to 2,737 objects (approximately 0.30\% of the initial sample), yielding a manageable set of candidates for subsequent human vetting with associated pipeline products. 
From this sample, visual inspection identifies 76 transient-like candidates as promising, which are selected for broader community dissemination and potential follow-up observations.

As a representative candidate used to illustrate the pipeline outputs, AT~2025ifl (pipeline object identifier: {\tt C202504181415355m003413}) is one of the selected candidates through the filtering described in Table~\ref{tab:vetting}. It is photometrically identified as a Type~Ia supernova located close to the nucleus of its host galaxy at $z=0.127$ \citep{NEDLVS_2023}. The pipeline products for this object are presented in Figure~\ref{fig:pipeline_product_FC} and Figure~\ref{fig:pipeline_product_LC}, showing the finding chart and the multi-band light curves, respectively.

\section{Pipeline Performance} \label{sec:pipeline_performance}

\subsection{Artifact rejection} \label{subsec:rejection_performance}

Effective artifact rejection is a critical component of any transient search pipeline. Difference imaging can produce a large number of spurious detections arising from imperfect image alignment and PSF mismatch, saturated sources, cosmic rays, and detector-related artifacts. 
In practice, this entails rejecting the vast majority of bogus detections while maintaining a high recovery rate for real astrophysical transients.

\begin{figure}[ht!]
\includegraphics[width=1.0\columnwidth]{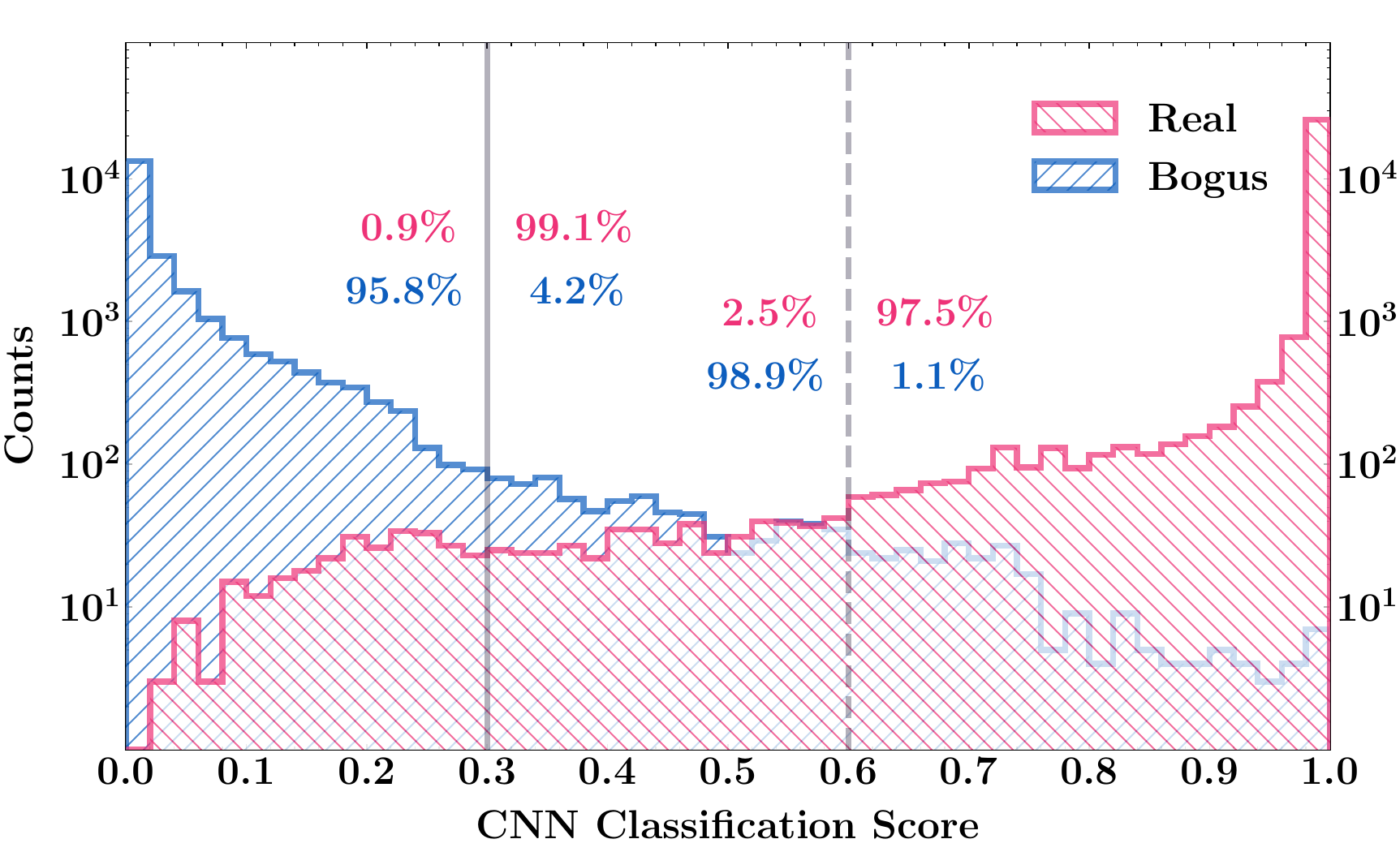}
\caption{
    Performance of the CNN real–bogus classifier. Distribution of CNN classification scores for detections in the test sample labeled as real (pink) and bogus (blue).
    Vertical lines mark two representative operating points of the classifier: the default hot threshold at ${\tt CNN\_SCORE\_THRESH}=0.3$ (solid line), used for alert generation, and an optional cold, purity-favoring threshold at ${\tt CNN\_SCORE\_THRESH}=0.6$ (dashed line). 
    The annotated percentages indicate the fractions of real and bogus candidates retained or rejected at each threshold.  \label{fig:pipeline_cnn}
}
\end{figure}

In our pipeline, artifact rejection is primarily achieved through a CNN real--bogus classifier, which serves as the first stage of transient filtering. By suppressing obvious artifacts at an early stage, the CNN significantly reduces the number of spurious detections passed to downstream candidate filtering and inspection. 
Using the DECam test sample described in Section~\ref{subsubsec:real_bogus}, we evaluate the performance of the CNN real--bogus classifier, as illustrated by the score distributions in Figure~\ref{fig:pipeline_cnn}. 

At the default hot threshold of ${\tt CNN\_SCORE\_THRESH}=0.3$, the classifier retains $\sim98.9\%$ of real detections while rejecting $\sim95.8\%$ of bogus detections. This configuration effectively suppresses the majority of obvious artifacts while maintaining a high recovery rate for real astrophysical transients, and is therefore adopted as the default setting for alert generation.
For applications that prioritize higher-confidence candidates and low human vetting overhead, the pipeline can optionally be configured with a more conservative cold threshold of ${\tt CNN\_SCORE\_THRESH}=0.6$, which we recommend as a representative high-purity operating point. Under this setting, $\sim98.9\%$ of bogus detections are rejected, while $\sim97.5\%$ of real detections are retained. This more aggressive filtering further reduces the number of spurious candidates passed to downstream inspection, at the expense of a modest loss of real detections.


\subsection{Computing performance} \label{subsec:conputing_performance}

\begin{figure*}[ht!]
\plotone{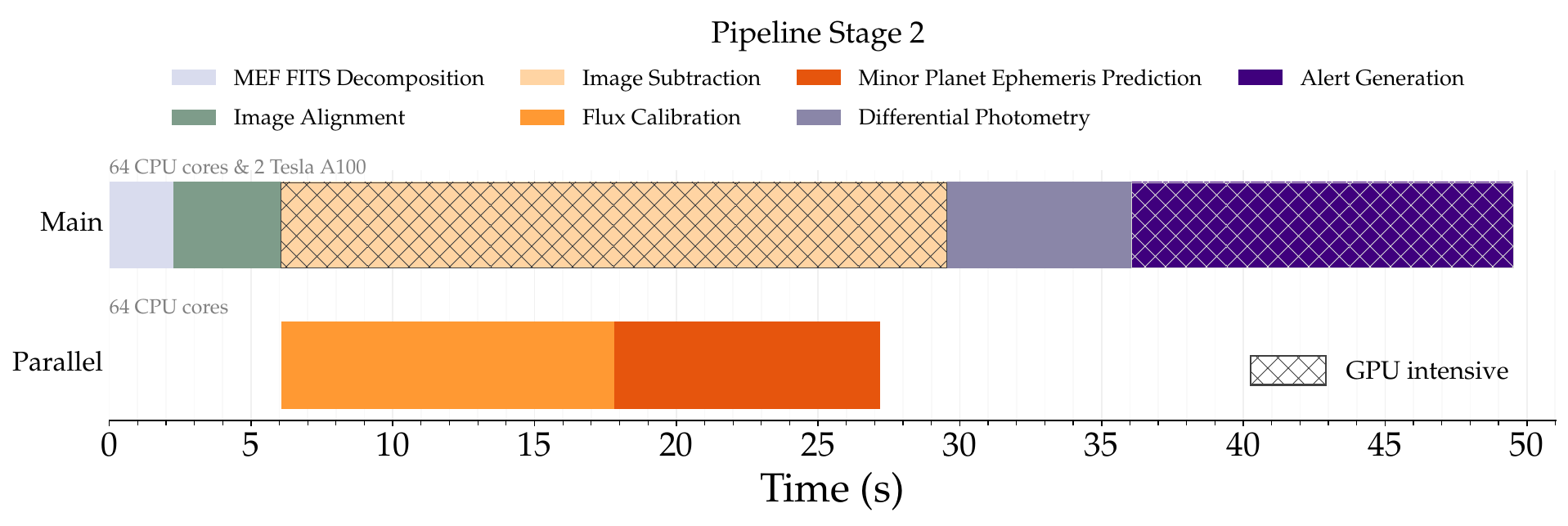}
\caption{Computing performance of the Stage~2 pipeline. Timeline of the main and parallel components in Stage~2 for processing a single DECam exposure. The main pipeline (top) runs with 64 CPU cores and two NVIDIA Tesla A100 GPUs on PSC/Vera, and includes MEF (multi-extension FITS) decomposition, image alignment, image subtraction, differential photometry, and alert generation. In parallel (bottom), flux calibration and minor-planet ephemeris prediction are executed concurrently on 64 CPU cores, overlapping in time with the main pipeline execution. Hatched blocks indicate GPU-intensive tasks. Under this configuration, the total wall-clock time for Stage~2 is approximately 50~s per exposure.\label{fig:pipeline_stage2}}
\end{figure*}


We evaluate the computing performance of the pipeline using real observations from the DESIRT 2025A program. Specifically, we randomly select five DECam observing nights in 2025A, comprising a total of 15 DECam exposures of a single pointing that covers the transient AT2025ifl. These data are processed with the full pipeline on the Vera cluster at the Pittsburgh Supercomputing Center (PSC).
The performance measurements reported below are based on these runs. Since data processing for different pointings with little or no spatial overlap is largely independent, using a single pointing provides a representative assessment of the pipeline performance for routine survey operations. The test follows the parameter configurations of DESIRT 2025A described in Section~\ref{sec:example_pipeline_products}.

The computational cost of each pipeline stage, as characterized by our performance tests on PSC/Vera, is summarized below.
\begin{itemize}
  \item \textbf{Stage~0 (pre-observation preparations):}  
  Stage~0 primarily involves the preparation of reference catalogs and the retrieval of template images. These operations are not computationally intensive and are largely constrained by network throughput. As a reference, compiling the reference catalogs by combining the Legacy Survey, PS1, and \textit{Gaia} data requires approximately $\sim6$~min per pointing on PSC/Vera. For template preparation, downloading a DECam exposure with calibrated data products (\texttt{ooi} and \texttt{ood}) from the NOIRLab Astro Data Archive requires approximately $\sim46$~s on our cluster. 

  Operationally, Stage~0 is executed after the observing plan is finalized but prior to data acquisition, with a typical lead time ranging from hours to days. This window is generally sufficient for completing Stage~0 tasks in advance of observations. Moreover, since Stage~0 is executed on a \textit{scheduled-instance} basis (defined by pointing and filter), its products can be reused for subsequent visits to the same field and filter. As a result, the cost of Stage~0 is amortized over repeated visits and Stage~0 does not typically set the processing latency of the pipeline.

  \item \textbf{Stage~1 (science image ingestion):} Stage~1 handles the ingestion of science images. Similar to Stage~0, this step is limited by network throughput rather than computation, with a per-exposure download time comparable to that of template retrieval (typically $\sim46$~s per exposure on PSC/Vera).

  \item \textbf{Stage~2 (image differencing and transient detection):} Stage~2 is the most computationally intensive component of the pipeline and sets the dominant per-exposure processing cost; it is therefore the primary focus of this section. Stage~2 performs image alignment, image subtraction, flux calibration, differential photometry, and alert generation, constituting the core image-processing and transient-detection workflow. 
  
  Figure~\ref{fig:pipeline_stage2} illustrates the execution timeline of Stage~2 for processing a single DECam exposure on PSC/Vera, averaged over the DECam test set. youThe main pipeline runs with 64 CPU cores and two NVIDIA Tesla A100 GPUs, while auxiliary tasks are executed concurrently on 64 CPU cores. The resulting total wall-clock time for Stage~2 is approximately $\sim50$~s per DECam exposure, demonstrating that the pipeline can keep pace with the typical DECam time-domain survey cadence.
  
  Consecutive DECam exposures are typically separated by at least $\sim1$~min, setting the practical requirement for per-exposure processing latency. The computing resource allocation for Stage~2 is designed to meet this requirement. Both GPU-intensive components—SFFT-based image differencing and CNN-based alert generation—operate on a CCD-by-CCD basis. For a single DECam exposure, each GPU processes a queue of approximately $\sim30$ CCD images sequentially. Owing to the high throughput of modern GPUs for these tasks, allocating a modest number of two GPUs is sufficient to process a full exposure within the inter-exposure interval, without accumulating processing backlogs during routine survey operations.

  Finally, we note that while the image processing components of Stage~2 exhibit stable performance across different fields, the runtime of photometry and alert generation can vary with image depth and field selection. The measurements reported here should therefore be interpreted as representative of typical DESIRT survey conditions.

  \item \textbf{Stage~3 (candidate filtering and pipeline product generation):} Stage~3 is usually executed after a set of DECam observations, for example on a nightly basis. Its computational cost is inherently workload-dependent, varying with the candidate filtering configuration and the number of DECam exposures associated with each selected candidate. In our test dataset, 132 candidates pass the regular pipeline selection criteria and have at least three detections (see Table~\ref{tab:vetting}). For this sample, Stage~3 requires approximately $\sim3$~s per selected candidate to generate the final pipeline products. In practice, the total cost of Stage~3 depends on the specific candidate filtering criteria and workload, but does not typically dominate the end-to-end processing latency of the pipeline.
\end{itemize}

\section{summary} \label{sec:summary}

We have presented a rapid transient detection pipeline designed to support a broad range of DECam time-domain surveys. The pipeline processes calibrated DECam imaging products provided by the DECam Community Pipeline, and delivers transient alerts and light-curve products on timescales suitable for (near-)real-time data analysis and rapid follow-up observations. It serves as the transient discovery engine for multiple DECam long-term survey programs, including GW-MMADS for gravitational-wave follow-up and the DESIRT intermediate-redshift extragalactic transient survey. 

The pipeline is organized into four stages. 
Stage~0 is executed prior to data acquisition and makes use of the time window between observation planning and execution to prepare reference catalogs and retrieve or construct template images (Section~\ref{subsubsec:prepsteps}). These preparatory tasks are designed to be executed only once for a given observing plan. 
Stage~1 handles the retrieval of calibrated DECam science images from the NOIRLab Astro Data Archive (Section~\ref{subsubsec:stage1}). 
Stage~2 constitutes the core of the pipeline, performing image differencing using the GPU-accelerated SFFT algorithm, followed by differential photometry and CNN-based real--bogus classification to suppress subtraction artifacts (Section~\ref{subsubsec:stage2}). 
Stage~3 applies transient candidate filtering and produces the final science-ready transient products, including finding charts and light curves, and is typically executed after a batch of exposures has been accumulated (Section~\ref{subsubsec:stage3}).

We demonstrate the application of the pipeline using DECam observations from the DESIRT survey obtained during the 2025A semester (Section~\ref{sec:example_pipeline_products}). This data set provides a survey-scale sample with a $\sim$half-year baseline, comprising thousands of exposures. 
We illustrate the candidate filtering performed by the pipeline, in which the combination of CNN-based real--bogus classification and dedicated pipeline selection criteria reduces the number of candidates requiring human vetting to a tractable level. A well-observed supernova from DESIRT 2025A, AT~2025ifl, is used to showcase the final science products generated by the pipeline.

We further evaluate the pipeline performance in Section~\ref{sec:pipeline_performance}. 
For the CNN-based real--bogus classification, the default configuration achieves a completeness of $>99\%$ for real astrophysical transients while rejecting $\sim$96\% of subtraction artifacts. A more purity-favored configuration yields a higher artifact rejection rate of $\sim$99\%, at the expense of a slightly lower completeness of $\sim$97.5\%.
In terms of computational performance, Stage~2, which performs image differencing and transient detection, is the most computationally intensive component of the pipeline and sets the overall per-exposure latency. 
With GPU acceleration, the pipeline achieves a per-exposure processing time of approximately 50~s on the PSC/Vera cluster using a modest allocation of computing resources, appropriate for time-critical DECam time-domain surveys.

\clearpage
\begin{acknowledgments}

We thank T. Zhang, D. Sand, C. Ransome, X. Li, Z. Xiong, and T. Sun for helpful discussions and suggestions.
LH, TC, AP are supported by NSF Grant No. 2308193. This work used resources on the Vera Cluster at the Pittsburgh Supercomputing Center (PSC). We thank the PSC staff for help with setting up our software on the Vera Cluster. 

This work used data obtained with the Dark Energy Camera (DECam), which was constructed by the Dark Energy Survey (DES) collaboration.
Funding for the DES Projects has been provided by the US Department of Energy, the US National Science Foundation, the Ministry of Science and Education of Spain, the Science and Technology Facilities Council of the United Kingdom, the Higher Education Funding Council for England, the National Center for Supercomputing Applications at the University of Illinois at Urbana-Champaign, the Kavli Institute for Cosmological Physics at the University of Chicago, Center for Cosmology and Astro-Particle Physics at the Ohio State University, the Mitchell Institute for Fundamental Physics and Astronomy at Texas A\&M University, Financiadora de Estudos e Projetos, Fundação Carlos Chagas Filho de Amparo à Pesquisa do Estado do Rio de Janeiro, Conselho Nacional de Desenvolvimento Científico e Tecnológico and the Ministério da Ciência, Tecnologia e Inovação, the Deutsche Forschungsgemeinschaft and the Collaborating Institutions in the Dark Energy Survey.

\end{acknowledgments}

\vspace{5mm}
\facilities{CTIO:4m}


\software{
    Astropy \citep{astropy13,astropy18,astropy22}, 
    SciPy \citep{2020SciPy-NMeth},
    Numpy \citep{2020NumPy-Array},
    Matplotlib \citep{Matplotlib},
    CuPy \citep{CuPy},
    Scikit-image \citep{skimage},
    SourceExtractor \citep{SExtractor},
    SWarp \citep{SWarp},
    SFFT \citep{sfft_zenodo},
    healpy \citep{Zonca2019},
    sklearn \citep{sklearn_2011}
}

\bibliography{PASPsample631}{}
\bibliographystyle{aasjournal}



\end{document}